\documentclass[11pt]{article}

\usepackage[T1]{fontenc}
\usepackage{lmodern}
\usepackage[margin=1in]{geometry}
\usepackage{amsmath,amsfonts,amssymb}
\usepackage{microtype}

\numberwithin{equation}{section}
\def\be{\begin{equation}}
\def\ee{\end{equation}}
\def\bq{\begin{eqnarray}}
\def\eq{\end{eqnarray}}
\def\beq{\begin{eqnarray*}}
\def\eeq{\end{eqnarray*}}

\begin{document}

\title{Asymptotic analysis of varying light speed cosmologies}
\author{Dimitrios Trachilis\\
College of Engineering and Technology,\\
American University of the Middle East, Egaila 54200, Kuwait\\
\texttt{Dimitrios.Trachilis@aum.edu.kw}}
\date{}

\maketitle
\begin{abstract}
\noindent We investigate potential singularities in isotropic cosmological models featuring a time-varying speed of light and the gravitational constant. The governing field equations generally simplify into two-dimensional systems, subject to analysis through dynamical systems techniques within phase space and the application of asymptotic splitting methods. In the broader context, our findings reveal the existence of initially expanding closed models that undergo subsequent recollapse towards a future singularity, and open universes exhibiting perpetual expansion into the future. The specific characteristics of these singularities are also expounded upon.
\end{abstract}


\section{Introduction}

Varying Speed of Light (VSL) cosmologies have garnered significant attention as alternatives to cosmological inflation, offering an alternative framework for addressing the challenges posed by the standard model (refer to \cite{mag} and associated references for further details). Rather than embracing the inflationary paradigm positing a phase of superluminal expansion in the early universe, these cosmologies postulate an accelerated speed of light during this epoch. In such cosmological frameworks, the conventional enigmas encountered in standard early universe cosmology are circumvented, albeit at the expense of violating the general covariance of the underlying gravitational theory. This response refrains from delving into a discourse on the foundational aspects of VSL theories and the conceptual intricacies stemming from the variable nature of light speed (see \cite{mag} for an in-depth discussion).

In this paper, we present initial findings from \cite{mi-co} involving the application of the asymptotic splittings method, as introduced in \cite{CB}, to analyze singularities that may manifest in Variable Speed of Light (VSL) cosmologies. Specifically, our focus is directed towards the VSL model proposed in \cite{alb-mag} and further elucidated in \cite{bar-mag}. A noteworthy feature of this model lies in the assumption of minimal coupling within the framework of Einstein's equations. This assumption posits that a time-variable speed of light (c) should not induce alterations in the curvature terms within Einstein's equations in the cosmological context; thus, the foundational equations of Einstein persist. As noted by Barrow \cite{go}, changes in the speed of light occur within the local Lorentzian frames associated with cosmological expansion, representing a special-relativistic effect. Consequently, the resulting theory lacks covariance, necessitating the adoption of a specific time coordinate. Opting for comoving proper time as this specific choice, the Friedman equations maintain their structure while accommodating variations in both c(t) and G(t).

The plan of this paper is as follows. In the next two sections, we use the Friedman equations with a power representation for the speed of light to derive an autonomous dynamical system in suitable variables. Then, we apply the method of asymptotic splittings to study the evolution of this model. We show that there exist asymptotic solutions of the unknown variables near the initial singularity. In sections 4-5, apart from the speed of light, we consider an additional power representation of $G$ and apply the same asymptotic method to find the admissible asymptotic solutions of this generalized model. In section 6, we connect the general solutions found from the method of asymptotic splittings to the flatness problem described in \cite{B1999} and \cite{B2003}.

We start with the Friedman equations,
\begin{eqnarray}
(\frac{\dot{a}}{a})^{2} &=& -\frac{kc^2}{a^2}+\frac{8\pi G}{3}\rho ,
\label{Friedman1}\\
\frac{\ddot{a}}{a} &=& -\frac{8\pi G}{6}(3\gamma -2)\rho ,
\label{Friedman2}
\end{eqnarray}
where the equation of state has the form,
\be
p=(\gamma -1)\rho c^2,
\label{eqstate}
\ee
with $0<\gamma \leq 2$. Here $a=a(t)$ is the scale factor, $p$ is the fluid pressure, $\rho$ is the fluid density, and $k$ is the constant curvature normalized to $k=0, +1$ or $-1$, for the flat, closed, or open space, respectively.

We consider the basic Friedman cosmological equations (\ref{Friedman1}) and (\ref{Friedman2}) for varying speed of light $c$ and for both constant and varying Newtonian gravitational $G$ to derive the associated conservation law of the theory. We show that these equations form an autonomous dynamical system in suitable variables. Further, we apply the method of asymptotic splittings given in \cite{CB} to examine the early evolution of the varying speed of light cosmology. In particular, we study separately $2+1$ cases with respect to the varying $c$ and $G$, namely that the subcases of varying $c$ with constant $G$, varying $c$ and $G$, and constant $c$ with varying $G$. The last subcase seems not to belong to a varying speed of light cosmology, but in fact it provides an equivalent problem. For every previous representation of $c$ and $G$ we find valid asymptotic solutions near the initial singularity $t_0$.

\section{Reduction to two dimensions with constant $G$}
We assume that the varying speed of light $c$ has a representation of the form,
\be
c=c_{0}a^{n}, \quad n\in \mathbb {R},
\label{cseries}
\ee
the system of units of which satisfies $8\pi G=1={c_0}^2$. Differentiating eq. (\ref{Friedman1}) and using eq. (\ref{Friedman2}), we find the conservation law, 
\be
\dot{\rho}+3\gamma \rho \frac{\dot{a}}{a}=\frac{3k}{4\pi G}\frac{c\dot{c}}{a^2}.
\label{conlaw}
\ee
Obviously, if $\dot{c}=0$, that is $n=0$, we obtain the same equations as in general relativity because the right-hand side of (\ref{conlaw}) vanishes.
Setting
\begin{eqnarray}
x &=& \frac{1}{a},
\label{eqx}\\
H &=& \frac{\dot{a}}{a},
\label{eqH}
\end{eqnarray}
the system (\ref{Friedman1}), (\ref{Friedman2}), and (\ref{conlaw}) implies,
\begin{eqnarray}
\dot{x} &=& -xH,
\label{dotx}\\
\dot{H} &=& -H^{2}-\frac{3\gamma -2}{6}\rho,
\label{dotH}\\
\dot{\rho} &=& -3\gamma \rho H + 6kn x^{2-2n}H,
\label{dotrho}\\
\end{eqnarray}
together with the constraint,
\be
H^2 + kx^{2-2n} = \frac{1}{3}\rho.
\label{constraint}
\ee
Using the last equation, we can eliminate $x$ or $\rho$. If we eliminate $x$, the remaining $2$-dimensional system takes the form,
\begin{eqnarray}
\dot{H} &=& -H^{2}-\frac{3\gamma -2}{6}\rho,\nonumber\\
\dot{\rho} &=& -(3\gamma -2n) \rho H - 6nH^3.
\label{system10}
\end{eqnarray}

\section{Asymptotic forms of $H$ and $\rho$ with constant $G$}
In this section we will apply the method of asymptotic splittings \cite{CB} (see also \cite{ckkt},\cite{kolionis1}) to the system (\ref{system10}) in order to find the dominant features on approach to the initial singularity, taken at $t=0$. In fact, the singularity can be replaced by any $t=t_0$ using the variable $\tau =t-t_0$. System (\ref{system10}) can be written as an autonomous system of the form, Below we are interested in finding asymptotic solutions that correspond to varying light speed cosmologies, thus we consider only the $n \neq 0$ cases. We start with,
\be
{\bf{\dot{x}}}={\bf {f}}({\bf{x}}), \quad {\bf {x}}=(H, \rho),
\label{autonomous}
\ee
and can be decomposed asymptotically into a dominant part and another, subdominant part
\be
{\bf {f}}({\bf{x}})={\bf {f}}^{(0)}_{i}({\bf{x}}) + {\bf {f}}^{(sub)}_{i}(\bf{x}),
\label{splitting}
\ee
in $9$ different ways $(i=1, 2, 3, \cdots, 9)$. A solution to the system (\ref{system10}), which is called \textit {dominant} near the singularity, is of the form,
\be
{\bf{x}}{(t)}={\bf{a}} t^{\bf{p}}=(\alpha t^{p}, \beta t^{q}),
\label{solution_form}
\ee
where ${\bf {a}}=(\alpha, \beta) \in \mathbb{C}^2$ and ${\bf {p}}=(p, q) \in \mathbb{Q}^2$. The pair $({\bf {a}}, {\bf {p}})$ is called a \textit {dominant balance} of the vector field ${\bf {f}}({\bf{x}})$ if it corresponds to a dominant solution of the system (\ref{system10}) near the singularity.

The first possible asymptotic decomposition is
\be
{\bf {f}}({\bf{x}})={\bf {f}}^{(0)}_{1}({\bf{x}}) + {\bf {f}}^{(sub)}_{1}(\bf{x}),
\label{first_splitting}
\ee
with dominant part,
\be
{\bf {f}}^{(0)}_{1}({\bf{x}})=\Big( -H^{2}, -(3\gamma -2n)\rho H \Big),
\label{first_splitting_dom}
\ee
and subdominant part,
\be
{\bf {f}}^{(sub)}_{1}({\bf{x}})=\Big( -\frac{3\gamma -2}{6}\rho, -6nH^{3} \Big),
\label{first_splitting_subdom}
\ee
Substituting
\be
H=\alpha t^{p} \quad \text{and} \quad \rho =\beta t^{q}
\label{H_rho}
\ee
into the dominant system ${\bf{\dot{x}}}={\bf {f}}^{(0)}_{1}({\bf{x}})$ we evaluate the dominant balance $({\bf {a}}, {\bf {p}})$. In particular, we find that,
\be
({\bf {a}}, {\bf {p}})=\Big((1,0), (-1,q) \Big), \quad q \leq -1.
\label{first_balance}
\ee
Further, we need to determine whether the term (\ref{first_splitting_subdom}) is indeed subdominant by evaluating the limit of this term divided by $t^{{\bf {p}}-1}$ when $t \rightarrow 0$. For values of $q \leq 2$, we find,
\be
\dfrac{{\bf {f}}^{(sub)}_{1}({\bf{x}})}{t^{{\bf {p}}-1}} \rightarrow (0,0), \quad \text{when} \quad t \rightarrow 0,
\label{first_limit_sub}
\ee
and therefore the decomposition (\ref{first_splitting}) is acceptable asymptotically. 

Next, we evaluate the Kovalevskaya matrix, given by,
\be
\mathcal{K}=\it {D} {\bf {f}}^{(0)}_{1}({\bf{a}}) - \text{diag}({\bf{p}}),
\label{Kov}
\ee
and we find that there is always one eigenvalue $\lambda =-1$ that corresponds to an arbitrary constant of the dominant behaviour of the solution given by the \textit {Fuchsian series expansions},
\be
H=\sum_{j=0}^{\infty}c_{j1}t^{p-j}, \quad \rho =\sum_{j=0}^{\infty}c_{j2}t^{q-j},
\label{Fuschian}
\ee
where $c_{01}=\alpha =1$ and $c_{02}=\beta =0$. Substituting the series (\ref{Fuschian}) into the original system (\ref{system10}), we find that there is no solution for $H$ and $\rho$ unless $n=0$. This is because of
our final test for admission of the solutions, namely that, the Fredholm's alternative,
\be
{\bf {v}}^{T}(\mathcal{K}-j\mathcal{I}){\bf {c}}_j={\bf {0}},
\label{Fredholm}
\ee
where ${\bf {v}}$ denote the eigenvectors of $\mathcal{K}$ and $\mathcal{I}$ is the identity matrix. In particular, in our first decomposition, the expression (\ref{Fredholm}) is true only for $n=0$. Therefore, the first decomposition does not give any valid asymptotic solution of the initial system around the singularity.

The second asymptotic decomposition is
\be
{\bf {f}}({\bf{x}})={\bf {f}}^{(0)}_{2}({\bf{x}}) + {\bf {f}}^{(sub)}_{2}(\bf{x}),
\label{second_splitting}
\ee
with dominant part,
\be
{\bf {f}}^{(0)}_{2}({\bf{x}})=\Big( -H^{2}, -6nH^{3} \Big),
\label{second_splitting_dom}
\ee
and subdominant part,
\be
{\bf {f}}^{(sub)}_{2}({\bf{x}})=\Big( -\frac{3\gamma -2}{6}\rho, -(3\gamma -2n)\rho H \Big),
\label{second_splitting_subdom}
\ee
Substituting $H=\alpha t^{p}$ and $\rho =\beta t^{q}$ into the second dominant system, we find the dominant balance,
\be
({\bf {a}}, {\bf {p}})=\Big((1,3n), (-1,-2) \Big).
\label{second_balance}
\ee
The corresponding subdominant term (\ref{second_splitting_subdom}) satisfies
\be
\dfrac{{\bf {f}}^{(sub)}_{2}({\bf{x}})}{t^{{\bf {p}}-1}}=\Big(\frac{2-3\gamma}{2}n,3n(2n-3\gamma)\Big),
\label{second_sub}
\ee
and therefore approaches $(0,0)$ if and only if $(\gamma =\frac{2}{3}, n=1)$ or $(\gamma \in \mathbb{R}, n=0)$. In principle, we can accept only the first pair. However, the expression (\ref{Fredholm}) does not give any acceptable solution for the combination $\gamma =\frac{2}{3}$ and $n=1$. Thus the second decomposition does not lead to a fully acceptable solution as well.

The third possible asymptotic decomposition is
\be
{\bf {f}}({\bf{x}})={\bf {f}}^{(0)}_{3}({\bf{x}}) + {\bf {f}}^{(sub)}_{3}(\bf{x}).
\label{third_splitting}
\ee
The dominant part is,
\be
{\bf {f}}^{(0)}_{3}({\bf{x}})=\Big( -H^{2},-(3\gamma -2n)\rho H -6nH^{3} \Big),
\label{third_splitting_dom}
\ee
and the subdominant part is,
\be
{\bf {f}}^{(sub)}_{3}({\bf{x}})=\Big( -\frac{3\gamma -2}{6}\rho, 0 \Big).
\label{third_splitting_subdom}
\ee
Substituting (\ref{H_rho}) into the third dominant system, we find,
\be
({\bf {a}}, {\bf {p}})=\Big((1,0), (-1,-2) \Big) \quad \text{and} \quad ({\bf {a}}, {\bf {p}})=\Big((1,3), (-1,-2) \Big).
\label{third_balance}
\ee
In case that $({\bf {a}}, {\bf {p}})=\Big((1,0), (-1,-2) \Big)$, we find no asymptotic solutions with $n \neq 0$, but in case that $({\bf {a}}, {\bf {p}})=\Big((1,3), (-1,-2) \Big)$, we obtain the special solution,
\be
H=t^{-1}, \quad \rho =3t^{-2}, \quad \text{where} \quad k=0, \gamma =\frac{2}{3}, n=-\frac{1}{2}.
\label{third_balance_sol3}
\ee
In addition, the corresponding subdominant term (\ref{third_splitting_subdom}) satisfies
\be
\dfrac{{\bf {f}}^{(sub)}_{3}({\bf{x}})}{t^{{\bf {p}}-1}} \rightarrow (0,0),
\label{third_sub}
\ee
when $t \rightarrow 0$, in this case.

The next three possible asymptotic decompositions,
\be
{\bf {f}}({\bf{x}})={\bf {f}}^{(0)}_{j}({\bf{x}}) + {\bf {f}}^{(sub)}_{j}({\bf{x}}), \quad j=4,5,6,
\label{456_splitting}
\ee
for which,
\begin{eqnarray}
{\bf {f}}^{(0)}_{4}({\bf{x}})&=&\Big( -\frac{3\gamma -2}{6}\rho,-(3\gamma -2n)\rho H \Big),
\label{fourth_splitting_dom} \\
{\bf {f}}^{(0)}_{5}({\bf{x}})&=&\Big( -\frac{3\gamma -2}{6}\rho,-6nH^3 \Big),
\label{fifth_splitting_dom} \\
{\bf {f}}^{(0)}_{6}({\bf{x}})&=&\Big( -\frac{3\gamma -2}{6}\rho,-(3\gamma -2n)\rho H -6nH^3 \Big),
\label{sixth_splitting_dom}
\end{eqnarray}
with corresponding subdominant parts,
\begin{eqnarray}
{\bf {f}}^{(sub)}_{4}({\bf{x}})&=&\Big( -H^2, -6nH^3 \Big),
\label{fourth_splitting_subdom} \\
{\bf {f}}^{(sub)}_{5}({\bf{x}})&=&\Big( -H^2, -(3\gamma -2n)\rho H \Big),
\label{fifth_splitting_subdom} \\
{\bf {f}}^{(sub)}_{6}({\bf{x}})&=&\Big( -H^2, 0 \Big),
\label{sixth_splitting_subdom}
\end{eqnarray}
give no asymptotic solution due to the fact that (\ref{fourth_splitting_subdom})-(\ref{sixth_splitting_subdom}) do not satisfy the conditions
\be
\dfrac{{\bf {f}}^{(sub)}_{j}({\bf{x}})}{t^{{\bf {p}}-1}} \rightarrow (0,0), \quad j=4,5,6, \quad \text{when} \quad t \rightarrow 0.
\label{456_sub}
\ee

We apply the method of asymptotic splittings for the remaining three asymptotic decompositions, the last of which is the \textit{all-terms dominant} decomposition. We have,
\be
{\bf {f}}({\bf{x}})={\bf {f}}^{(0)}_{j}({\bf{x}}) + {\bf {f}}^{(sub)}_{j}({\bf{x}}), \quad j=7,8,9,
\label{789_splitting}
\ee
where,
\begin{eqnarray}
{\bf {f}}^{(0)}_{7}({\bf{x}})&=&\Big( -H^{2}-\frac{3\gamma -2}{6}\rho,-(3\gamma -2n)\rho H \Big),
\label{seventh_splitting_dom} \\
{\bf {f}}^{(0)}_{8}({\bf{x}})&=&\Big( -H^{2}-\frac{3\gamma -2}{6}\rho,-6nH^3 \Big),
\label{eighth_splitting_dom} \\
{\bf {f}}^{(0)}_{9}({\bf{x}})&=&\Big( -H^{2}-\frac{3\gamma -2}{6}\rho,-(3\gamma -2n)\rho H -6nH^3 \Big),
\label{ninth_splitting_dom}
\end{eqnarray}
and,
\begin{eqnarray}
{\bf {f}}^{(sub)}_{7}({\bf{x}})&=&\Big( 0, -6nH^3 \Big),
\label{seventh_splitting_subdom} \\
{\bf {f}}^{(sub)}_{8}({\bf{x}})&=&\Big( 0, -(3\gamma -2n)\rho H \Big),
\label{eighth_splitting_subdom} \\
{\bf {f}}^{(sub)}_{9}({\bf{x}})&=&\Big( 0, 0 \Big),
\label{ninth_splitting_subdom}
\end{eqnarray}
In every previous decomposition (\ref{seventh_splitting_dom})-(\ref{ninth_splitting_dom}), we find that $p=-1$ and $q=-2$. However, the possible values of $\alpha$ and $\beta$ that are accepted from the decomposed systems vary, and give the following six special asymptotic solutions for $H$ and $\rho$ in (\ref{system10}),
\begin{eqnarray}
H&=&\frac{1}{1-c}t^{-1}, \quad \rho =\frac{3c}{(1-c)^2}t^{-2}, \quad \text{where} \quad k \neq 0, \gamma =0, \quad \text{and \quad} n=c \quad \text{satisfies} \nonumber \\
&& -18c^{3}+89c^{2}-106c-1=0, 
\label{ninth_balance_sol4} \\
H&=&\frac{c-1}{108c}t^{-1}, \quad \rho =\frac{(c-1)^3}{419904c^2}t^{-2}, \quad \text{where} \quad (k=0, n=c=109, \gamma = \frac{218}{3}), \quad \text{or} \nonumber\\
&&(k<0, n=c=-53 \pm 6\sqrt{78}, \gamma =\frac{2c}{3}),
\label{eight_balance_sol2} \\
H&=&\frac{1}{1-c}t^{-1}, \quad \rho =\frac{6c}{(1-c)^{2}(2-3\gamma)}t^{-2}, \quad \text{where} \quad k=0, \pm 1, \gamma \neq 0,\frac{2}{3}, n=c \neq 1,\nonumber \\ 
\quad &&\text{and} \quad (3\gamma -2)^{2}(c^{2}+34c+1)+36c(c-1)(9\gamma +2c-8)=0, 
\label{ninth_balance_sol89} \\
H&=&\frac{1}{1-c}t^{-1}, \quad \rho =\frac{3}{(1-c)^{2}}t^{-2}, \quad \text{where} \quad k=0, \gamma =\frac{2-2n}{3}, \quad n=c \neq 1,\nonumber \\ 
&&\text{and} \quad c^{3}-2c^{2}+19c+18=0,
\label{ninth_balance_sol10} \\
H&=&\frac{2}{3\gamma}t^{-1}, \quad \rho =\frac{4}{3{\gamma}^{2}}t^{-2}, \quad \text{where} \quad k=0, \gamma \neq 0,\frac{2}{3}, \quad n \neq 1, \nonumber\\ 
\quad &&\text{and} \quad {\gamma}^{2}(3\gamma -2)(3\gamma -7)+8(3\gamma +4n)(3\gamma -2n)=0,
\label{ninth_balance_sol567}
\end{eqnarray}
and the general asymptotic solution,
\begin{eqnarray}
H&=&t^{-1}+c_{11}t^{-2}+\cdots, \quad \rho =3t^{-2}+c_{12}t^{-3}+\cdots, \quad \text{where} \quad c_{12}=3c_{11}, \nonumber\\
&&\text{and} \quad k=0, \pm 1, \gamma =\frac{2}{3}, n=-\frac{1}{2}.
\label{ninth_balance_sol23}
\end{eqnarray}
Solution (\ref{ninth_balance_sol23}) contains two arbitrary constants, $c_{11}$ and one more that corresponds to the arbitrary position of the singularity (here without loss of generality, we consider $t_{0}=0$). Notice that the special solution (\ref{third_balance_sol3}) found in the third decomposition is just a special case of (\ref{ninth_balance_sol23}).
Our various series solutions found in all nine asymptotic decompositions satisfy the phase space of the initial system (\ref{system10}) that follows from Eq. (\ref{constraint}). In particular, for closed models, the phase portrait of (\ref{system10}) is given by the set $\big\{ (H,\rho) \in \mathbb{R}^2 : \rho \geq 0, \rho -3H^{2}>0  \big\}$, and for open models it is given by $\big\{ (H,\rho) \in \mathbb{R}^2 : \rho \geq 0, \rho -3H^{2}<0  \big\}$.

\section{Reduction to two dimensions with varying $G$}
In this section we assume that both functions $c=c(t)$ and $G=G(t)$ have a power-law dependence of the form,
\be
c=c_{0}a^{n}, \quad G=G_{0}a^{m}, \quad n,m\in \mathbb {R},
\label{cGseries}
\ee
where the system of units requires $8\pi G_0 = {c_0}^{2}=1$. In this general case, if we differentiate (\ref{Friedman1}) and use (\ref{Friedman2}), we obtain the generalized conservation law, 
\be
\dot{\rho}+3\gamma \rho \frac{\dot{a}}{a}=-\rho \frac{\dot{G}}{G}+\frac{3k}{4\pi G}\frac{c\dot{c}}{a^2}.
\label{general_conlaw}
\ee
Setting again $x=\frac{1}{a}$ and $H=\frac{\dot{a}}{a}$, our initial system (\ref{Friedman1}), (\ref{Friedman2}), together with (\ref{general_conlaw}), becomes,
\begin{eqnarray}
\dot{x} &=& -xH,
\label{gen_dotx}\\
\dot{H} &=& -H^{2}-\frac{3\gamma -2}{6}\rho x^{-m},
\label{gen_dotH}\\
\dot{\rho} &=& -(m+3\gamma) \rho H + 6kn x^{m+2-2n}H,
\label{gen_dotrho},
\end{eqnarray}
subject to the constraint,
\be
H^2 + kx^{2-2n} = \frac{1}{3}\rho x^{-m}.
\label{gen_constraint}
\ee
Using (\ref{gen_constraint}) we can only eliminate $\rho$ and get a two-dimensional system with respect to $(x, H)$. The system that we find is,
\begin{eqnarray}
\dot{x} &=& -xH,\nonumber\\
\dot{H} &=& -\frac{3\gamma}{2}H^{2} - \frac{3\gamma -2}{2}kx^{2-2n}.
\label{system13}
\end{eqnarray}
Notice that the value of $m$ is not present in (\ref{system13}). However, it must be present in the solution of $\rho$ due to the constraint (\ref{gen_constraint}).
\section{Asymptotic forms with varying $G$}
In this section we will apply the method of asymptotic splittings to the system (\ref{system13}) generated by the Friedman equations with a power-law representation of $c$ and $G$. There are three different decomposed systems to apply the method of asymptotic splittings, that is,
\be
{\bf {f}}_{gen}({\bf{x}})={\bf {f}}^{(0)}_{gen,j}({\bf{x}}) + {\bf {f}}^{(sub)}_{gen,j}({\bf{x}}), \quad j=1,2,3,
\label{gen_splitting}
\ee
where,
\begin{eqnarray}
{\bf {f}}^{(0)}_{gen,1}({\bf{x}})&=&\Big( -xH, -\frac{3\gamma}{2}H^2 \Big),
\label{gen_splitting_dom1} \\
{\bf {f}}^{(0)}_{gen,2}({\bf{x}})&=&\Big( -xH, -\frac{3\gamma -2}{2}kx^{2-2n} \Big),
\label{gen_splitting_dom2} \\
{\bf {f}}^{(0)}_{gen,3}({\bf{x}})&=&\Big( -xH, -\frac{3\gamma}{2}H^2 -\frac{3\gamma -2}{2}kx^{2-2n} \Big),
\label{gen_splitting_dom3}
\end{eqnarray}
and,
\begin{eqnarray}
{\bf {f}}^{(sub)}_{gen,1}({\bf{x}})&=&\Big( 0, -\frac{3\gamma -2}{2}kx^{2-2n} \Big),
\label{gen_splitting_subdom1} \\
{\bf {f}}^{(sub)}_{gen,2}({\bf{x}})&=&\Big( 0, -\frac{3\gamma}{2}H^2 \Big),
\label{gen_splitting_subdom2} \\
{\bf {f}}^{(sub)}_{gen,3}({\bf{x}})&=&\Big( 0, 0 \Big),
\label{gen_splitting_subdom3}
\end{eqnarray}
The second decomposition (\ref{gen_splitting_dom2}) does not give any asymptotic solution to the problem, because the \textit{compatibility condition} (\ref{Fredholm}) is not valid in this case. However, the first decomposition and the \textit{all-terms dominant} case provide the following admissible asymptotic solution of the system (\ref{system13}) near the initial singularity,
\begin{eqnarray}
H&=&\frac{2}{3\gamma}t^{-1}, \quad x =c_{11}t^{-\frac{2}{3\gamma}}, \quad \text{where} \quad c_{11} \neq 0, m \in \mathbb{R}, n<1,\nonumber \\ 
\quad &&\text{with} \quad (k=0, \gamma \neq 0) \quad \text{or} \quad (k=0, \pm 1, \gamma = \frac{2}{3}), 
\label{firstgen_balance_sol145678_thirdgen_balance_sol12}
\end{eqnarray}
In fact, solution (\ref{firstgen_balance_sol145678_thirdgen_balance_sol12}) appears on both first and \textit{all-terms dominant} decompositions. This solution for $H$ and $x$ belongs to the phase space of the system given in(\ref{system13}), which is the set $\big\{ (x,H) \in \mathbb{R}^2 : x \geq 0, H^{2}+kx^{2-2n}>0  \big\}$.

Taking further into account the constraint (\ref{gen_constraint}), which represents the first Friedman equation (\ref{Friedman1}), we can have the asymptotic solutions for $H$ and $\rho$. In particular, the solution found in (\ref{firstgen_balance_sol145678_thirdgen_balance_sol12}) imply,
\begin{eqnarray}
H&=&\frac{2}{3\gamma}t^{-1}, \quad \rho =\frac{4}{3{\gamma}^2}c_{11}^{m}t^{-2-\frac{2m}{3\gamma}}+3kc_{11}^{2+m-2n}t^{\frac{4-4n}{3\gamma}-\frac{2m}{3\gamma}}, \quad c_{11} \neq 0, m \in \mathbb{R}, n<1,\nonumber \\ 
\quad &&\text{with} \quad (k=0, \gamma \neq 0) \quad \text{or} \quad (k=0, \pm 1, \gamma = \frac{2}{3}),
\label{firstgen_balance_sol145678_thirdgen_balance_sol12_Hrho}
\end{eqnarray}   

It is interesting to mention that the analysis we presented for the varying $G$ case can also be used for the special case $(n=0, m \in \mathbb{R})$. Taking into account Einstein's field equation,
\be
R_{ij}-\frac{1}{2}g_{ij}R=\frac{8\pi G}{c^4}T_{ij}
\label{Einstein field eq}
\ee
and (\ref{cGseries}), the choice to keep $G$ constant as $c$ varies or to keep $c$ constant as $G$ varies can be considered equivalent. In both cases the ratio $\frac{8\pi G}{c^4}$ takes the form $a^{\nu}=a^{m-4n}$. Of course, the two Friedman equations (\ref{Friedman1}) and (\ref{Friedman2}) lead to a different physical meaning, because $c$ is multiplying the scale factor term $a^{-2}$ while $G$ is multiplying the density term $\rho$. Returning back to our solutions, if $n=0$ and $m \neq 0$, the right-hand sides of the Friedman equations, that we started with, do not depend only on the variation of $a$ and $\rho$, but they depend on the variation of $G$ as well, and it is clear that the solution (\ref{firstgen_balance_sol145678_thirdgen_balance_sol12}) will remain invariant under $n=0$ ($n$ can be any number less than $1$), while (\ref{firstgen_balance_sol145678_thirdgen_balance_sol12_Hrho}) will give,
\begin{eqnarray}
H&=&\frac{2}{3\gamma}t^{-1}, \quad \rho =\frac{4}{3{\gamma}^2}c_{11}^{m}t^{-2-\frac{2m}{3\gamma}}+3kc_{11}^{2+m}t^{\frac{4}{3\gamma}-\frac{2m}{3\gamma}}, \quad c_{11} \neq 0, m \neq 0, n=0,\nonumber \\ 
\quad &&\text{with} \quad (k=0, \gamma \neq 0) \quad \text{or} \quad (k=0, \pm 1, \gamma = \frac{2}{3}).
\label{firstgen_balance_sol145678_thirdgen_balance_sol12_Hrho_n=0}
\end{eqnarray}

\section{Asymptotic solutions to the flatness problem}

Starting with the generalized conservation law (\ref{general_conlaw}) with $c=c_{0}a^{n}$ and $G=G_{0}a^{m}$ we obtain,
\be
(a^{m+3\gamma}\rho)^{\cdot} = 6knc_{0}^{2}a^{2n+3\gamma -3}\dot{a}.
\label{general_conlaw_v2}
\ee
Integrating (\ref{general_conlaw_v2}), we get,
\be
\rho = \frac{6kna^{2n-2-m}}{2n+3\gamma -2} + Ma^{-3\gamma -m}, \quad \text{if} \quad 2n \neq 2-3\gamma,
\label{general_conlaw_v2_int1}
\ee
or
\be
\rho = 3k(2-3\gamma)a^{-3\gamma -m}\ln{a} + Ma^{-3\gamma -m}, \quad \text{if} \quad 2n = 2-3\gamma.
\label{general_conlaw_v2_int2}
\ee
In particular, taking into account the general asymptotic solutions (\ref{ninth_balance_sol23}) and (\ref{firstgen_balance_sol145678_thirdgen_balance_sol12_Hrho}), we are interested in choosing $\gamma =\frac{2}{3}$, because this is the only choice that allows $k$ to be $0$ or $\pm 1$. Then, (\ref{general_conlaw_v2_int1}) and (\ref{general_conlaw_v2_int2}) read,
\be
\rho = 3kna^{2n-2-m} + Ma^{-2-m}, \quad \text{if} \quad n \neq 0,
\label{general_conlaw_v2_int1_v2}
\ee
and
\be
\rho = Ma^{-2-m}, \quad \text{if} \quad n=0,
\label{general_conlaw_v2_int2_V2}
\ee
respectively. Next, returning to the first Friedman equation (\ref{Friedman1}) we have,
\be
H^{2}=\frac{M}{3}a^{-2}+(n-1)ka^{2(n-1)}, \quad \text{if} \quad n \neq 0,
\label{H_n not zero}
\ee
and
\be
H^{2}=(\frac{M}{3}-k)a^{-2}, \quad \text{if} \quad n=0,
\label{H_n equals zero}
\ee
Considering (\ref{H_n not zero}) at large $a$, it is clear that the first term dominates the constant curvature term whenever $n \leq 0$, while in (\ref{H_n equals zero}) $k$ does not play any important role as it can be neglected by specific values of the integrating constant $M$. Therefore, in both cases, the flatness problem can be solved if $n \leq 0$. Notice that the value of $m$ does not have any contribution to the results. This is because $G$ only multiplies $\rho$ in (\ref{Friedman1}) and their product eliminates the constant $m$.

Our expressions (\ref{ninth_balance_sol23}) and (\ref{firstgen_balance_sol145678_thirdgen_balance_sol12_Hrho}) for constant and varying $G$ respectively, accept solutions for which $n \leq 0$, so they can also be considered as asymptotic solutions to the flatness problem. In particular, (\ref{ninth_balance_sol23}) remains the same, but (\ref{firstgen_balance_sol145678_thirdgen_balance_sol12_Hrho}) and its special case (\ref{firstgen_balance_sol145678_thirdgen_balance_sol12_Hrho_n=0}) become,
\be
H=t^{-1}, \quad \rho =\frac{1}{3}c_{11}^{m}t^{-2-m}+3kc_{11}^{2+m-2n}t^{2-2n-2m}, \quad c_{11} \neq 0, m \in \mathbb{R}, n \leq 0,
\label{firstgen_balance_sol145678_thirdgen_balance_sol12_Hrho_flatness}
\ee 
and
\be
H=t^{-1}, \quad \rho =\frac{1}{3}c_{11}^{m}t^{-2-m}+3kc_{11}^{2+m}t^{2-m}, \quad c_{11} \neq 0, m \neq 0, n=0,
\label{firstgen_balance_sol145678_thirdgen_balance_sol12_Hrho_n=0_flatness}
\ee
respectively.

\section{Conclusion} 
Comparing the results for both constant and varying $G$ by taking only into account the asymptotic solutions that express a varying speed of light cosmological theory ($n \neq 0$), we find that, in both cases, the associated general solutions (\ref{ninth_balance_sol23}) and (\ref{firstgen_balance_sol145678_thirdgen_balance_sol12_Hrho}) are not invariant with respect to the time-symmetry $t \rightarrow{-t}$, $H \rightarrow{-H}$, $\rho \rightarrow{\rho}$. This is due to the form of the energy density $\rho$ in the solutions. In particular, in (\ref{ninth_balance_sol23}) the expression for $\rho$  contains odd powers of $t$. Also, in (\ref{firstgen_balance_sol145678_thirdgen_balance_sol12_Hrho}) there are infinitely many combinations of the parameters $m$, $n$, and $\gamma$ that do not respect the symmetry $\rho \rightarrow{\rho}$ (this is also true for (\ref{firstgen_balance_sol145678_thirdgen_balance_sol12_Hrho_n=0}) when $n=0$ and $m \neq 0$). This means that for early times the existence of singularity related to the dynamical systems (\ref{system10}) and (\ref{system13}) creates a time-asymmetry.

In this paper, we have considered the behavior of the VSL cosmological models near the initial singularity for both $G=$constant and varying $G=G(t)$. Taking into consideration various asymptotic conditions that each decomposed system must satisfy in order to have admissible solutions near the initial singularity, we have presented the form of the series solutions of the variables of the systems (\ref{system10}) and (\ref{system13}) that are accepted.

It turns out that in both cases of $G$, the Hubble parameter behaves as $t^{-1}$ near the initial singularity. By applying various asymptotic arguments, we were able to built solutions to the field equations in both systems (\ref{system10}) and (\ref{system13}) in the form of a Fuchsian formal series expansion which are consistent with all other constraints, and are asymptotically dominant to leading order. Thus, we conclude that the given solutions correspond to early time attractors of all such homogeneous and isotropic VSL cosmologies. The same solutions for $n \leq 0$ can also be used for the solution to the flatness problem.

\bibliographystyle{plain}
\bibliography{references}

\end{document}